\documentclass[aps,prl,twocolumn]{revtex4}

\def\la{\langle}
\def\ra{\rangle}

 \def\be{\begin{eqnarray}}
\def\ee{\end{eqnarray}}

\def\ij{\langle ij\rangle}

\usepackage{graphicx}

\begin{document}

\title{Kosterlitz-Thouless Universality in Dimer Models}
\author{Shailesh Chandrasekharan and Costas G. Strouthos}
\affiliation{
Department of Physics, Box 90305, Duke University,
Durham, North Carolina 27708, USA.}

\preprint{DUKE-TH-03-243}

\begin{abstract}
Using the monomer-dimer representation of strongly coupled $U(N)$ lattice 
gauge theories with staggered fermions, we study finite temperature chiral 
phase transitions in $(2+1)$ dimensions. A new cluster algorithm allows us 
to compute monomer-monomer and dimer-dimer correlations at zero monomer 
density (chiral limit) accurately on large lattices. This makes it possible 
to show convincingly, for the first time, that these models undergo a finite 
temperature phase transition which belongs to the 
Kosterlitz-Thouless universality class. We find that this universality class 
is unaffected even in the large $N$ limit. This shows that the mean field 
analysis often used in this limit breaks down in the critical region.
\end{abstract}

\maketitle

Computing quantities in Lattice QCD with massless quarks is notoriously 
difficult. Most known algorithms break down in the chiral limit. For this
reason questions related to the universality of chiral phase transitions are
among the many questions that remain unanswered. It is often 
difficult to compute critical exponents sufficiently accurately to 
rule out all possibilities except one.

The most useful simplification of lattice QCD occurs in the strong
coupling limit which retains much of the underlying physics of QCD
except for large lattice artifacts. In this limit spontaneous chiral 
symmetry breaking and its restoration due to finite temperature 
effects have been studied using large $N$ and large $d$ 
expansions \cite{Kaw81,Klu83,Dam84}. However, since these approaches are 
based on mean field analysis they cannot help in determining the 
universality of phase transitions. 

Interestingly lattice QCD with staggered fermions interacting through 
$U(N)$ gauge fields can be mapped into a monomer-dimer system in the 
strong coupling limit \cite{Ros84}. These models contain an exact 
$U(1)$ chiral symmetry, a remnant of the full chiral symmetry of QCD. 
When it was proposed, the monomer-dimer representation offered a new
approach to study strongly coupled gauge theories close to the chiral 
limit from first principles. Unfortunately, this dream has remained 
unfulfilled until now. As in the weak coupling regime, most numerical 
simulations of the monomer-dimer systems have suffered from critical 
slowing down close to the chiral limit and hence have only allowed 
calculations with limited accuracy \cite{Boy92}.

Recently, a cluster algorithm has been discovered to study these 
strongly coupled lattice gauge theories in the chiral limit \cite{Cha03}.
This allows precision calculations in the chiral limit for the first time.
As a first application of this new algorithm, in this article we study 
the finite temperature critical behavior in $(2+1)$ dimensions. In agreement
with expectations from universality, we find with very high precision the 
chiral phase transition to be in the same universality class as the 
Berezinski-Kosterlitz-Thouless (BKT) transition \cite{BKT}. Our results 
are comparable to other known high precision spin-model studies of this 
universality class.
We also show that the BKT transition persists even in the large $N$ 
limit, showing that the mean field analysis breaks down in the critical 
region.

Although our results will primarily be of interest to lattice gauge 
theorists, it is possible that they may have a wider audience.
We are basically studying monomer-dimer models and such models have a 
long history \cite{Fow37,Hei72}. They have been studied extensively in 
the context of statistical and condensed matter physics. In the sixties, 
these models attracted a lot of attention \cite{Kas63, Fis63} when it 
was shown that the Ising model can be rewritten as a dimer model 
\cite{Fis66}. In the late eighties, these models became fashionable again 
in the quantum version \cite{Rok88} as a promising approach to the famous 
Resonating-Valence-Bond (RVB) liquid phase \cite{And87}. More recently the 
quantum dimer approach has again gained momentum since it was shown that 
the RVB liquid phase is actually realized on a triangular lattice 
\cite{Moe01}. This has raised hope that dimer models may also yield a 
theoretical understanding of the spin-liquid phase \cite{Moe02}. Inspite
of these studies, as far as we know, a BKT transition has not been studied 
in the context of a dimer model.

The partition function of the model we study in this article is given by 
(\cite{Ros84})
\begin{equation}
Z = \int [dU] d\psi d\bar\psi\ \exp\left(-S[U,\psi,\bar\psi]\right),
\label{uNpf}
\end{equation}
where $[dU]$ is the Haar measure over $U(N)$ matrices and $d\psi$
$d\bar\psi$ specify Grassmann integration. At strong couplings, the 
Euclidean space action $S[U,\psi,\bar\psi]$ is given by the fermionic 
part
\begin{equation}
\label{fact}
- \sum_{\ij} \eta_{\ij}\Big[\bar\psi_iU_{\ij} \psi_j
- \bar\psi_j U^\dagger_{\ij} \psi_i\Big]
- m \sum_i \bar\psi_i\psi_i,
\end{equation}
where $i$ refers to the lattice site on a periodic three dimensional cubic 
lattice of size $L_x,L_y$ along the two spatial directions and size $L_t$ 
along the euclidean time direction, $\ij$ represents the bond connecting 
the nearest neighbor sites $i$ and $j$, $U_{\ij}$ are $N\times N$ unitary 
matrices associated with the bond $\ij$ and represent the gauge fields, 
$\psi_i$ is an $N$ component column vector and $\bar\psi_i$ is an $N$ 
component row vector made up of Grassmann variables and represent staggered 
fermion fields at the site $i$. We will assume that the gauge links satisfy 
periodic boundary conditions while the fermion fields satisfy either periodic 
or anti-periodic boundary conditions. The factors $\eta_{\ij}$ are the
well known staggered fermion phase factors that depend on the coordinates
$i$(or $j$). We will choose them to have the property that 
$\eta_{\ij}^2 = 1$ when $\ij$ is a spatial bond, and $\eta_{\ij}^2 = T$
when $\ij$ is a temporal bond. The exact form of $\eta_{\ij}$ is not 
important at strong couplings since only $\eta_{\ij}^2$ appears in the 
final partition function. 
The real parameter $T$ acts like a temperature. By working on asymmetric 
lattices with $L_t << L$  and allowing $T$ to vary continuously, one can 
study finite temperature phase transitions in strong coupling QCD 
\cite{Boy92}.

The partition function given in eq.(\ref{uNpf}) can be rewritten as a
monomer-dimer system given by
\begin{equation}
Z \;=\; \sum_{[n,b]}' \;\;\;\;\;
\prod_{\ij}\; (z_{\ij})^{b_{\ij}}\frac{(N-b_{\ij})!}{b_{\ij}! N!}\;\;\; 
\prod_i \frac{N!}{n_i!}\;m^{n_i},
\label{pf}
\end{equation}
and is discussed in detail in \cite{Ros84,Cha03}. Here $n_i=0,1,2,...,N$ 
refers to the number of monomers on the site $i$,  $b_{\ij}=0,1,2,...,N$ 
represents the number of dimers on the bond $\ij$, $m$ is the monomer weight, 
$z_{\ij}=\eta_{ij}^2$ are the dimer weights. Note that while spatial dimers
carry a weight $1$, temporal dimers carry a weight $T$. The sum is over 
all monomer-dimer configurations $[n,b]$ which are constrained such that 
the sum of the number of monomers at each site and the dimers that touch 
the site is always $N$ (the number of colors). The $'$ in the sum reminds 
us of this constraint. In this work we choose $L_x=L_y=L$.

Comparing eqs. (\ref{pf}) and (\ref{uNpf}) we learn that zero monomer 
density corresponds to the chiral limit. The bipartite nature of the 
lattice can be used to distinguish every site $j$ as either even or odd. 
If we define $\sigma_j = +1$ for an even site and
$\sigma_j = -1$ for an odd site, it is easy to show that when $m=0$ the 
action given in eq. (\ref{fact}) is invariant under $U(1)$ chiral 
transformations,
\begin{equation}
\psi(j) \rightarrow \mathrm{e}^{i\sigma_j \theta}\psi(j),\ \ 
\bar\psi(j) \rightarrow \bar\psi(j)\mathrm{e}^{i\sigma_j\theta}.
\end{equation}
The results of \cite{Cha03} show convincingly that this $U(1)$ 
chiral symmetry breaks spontaneously at zero temperatures in two 
spatial dimensions. In this article we study the finite temperature
critical behavior and show that it belongs to the BKT universality class 
as expected. For this purpose we compute the chiral susceptibility 
$\chi_s$ and the winding number susceptibility $\chi_W$, both of which 
are often computed in this context. The chiral susceptibility is equivalent
to the spin susceptibility of classical $XY$ models and is given by
\begin{equation}
\chi_s = \frac{1}{V} \sum_{i,j} \la \bar\psi_i\psi_i \bar\psi_j\psi_j \ra
= \frac{1}{V Z}\frac{\partial^2 Z}{(\partial m)^2}\Big|_{m=0}
\end{equation}
where $V = L^2 L_t$ is the lattice volume. It is easy to see that $\chi_s$
is the integrated monomer-monomer correlation function.
We define the winding number susceptibility by
\begin{equation}
\chi_W = \frac{\pi}{2} \la W_x^2 + W_y^2 \ra.
\end{equation}
where $W_x$ and $W_y$ are the winding numbers in the $x$ and $y$ directions 
respectively and are given by $W_x=(\sum_i \sigma_i b_{\ij})/L_x$,and 
$W_y=(\sum_i \sigma_i b_{\ij})/L_y$. In the definition of $W_x$ and $W_y$
the site $j$ is chosen such that the bond $\ij$ is along the positive 
$x$ and $y$ directions respectively.

In this study we fix $L_t$ and compute $\chi_s$ and $\chi_W$ as a 
function of $L$. There are striking predictions for the large $L$ 
behavior of these quantities if the phase transition belongs to the 
BKT universality class. If $T_c$ represents the 
critical temperature we expect 
\begin{equation}
\chi_s \propto \left\{\begin{array}{lc}
L^{2-\eta(T)} & T<T_c \cr
L^{1.75}\ [\log(L)]^{0.125} & T=T_c \cr
\mbox{constant} & T>T_c  \cr
\end{array}\right.
\end{equation}
and
\begin{equation}
\chi_W = \left\{\begin{array}{lc}
1/[2\eta(T)] & T<T_c \cr
[2 + 1/\log(L/L_0)] & T=T_c \cr
0 & T>T_c  \cr
\end{array}\right.
\end{equation}
in the large $L$ limit. At $T=T_c$ the next to leading order correction
is also shown since it can be important. The critical exponent $\eta(T)$ 
is expected to change continuously
with temperature but remain in the range $0\leq \eta(T) < 0.25$ assuming
the monomers carry a unit of $U(1)$ charge. 
These predictions have been discussed in \cite{BKT,Ken95,Nel77,Pol87} 
and have been used in \cite{Har98,Cha02} earlier to demonstrate 
BKT behavior. In order to confirm these predictions we 
have computed $\chi_s$ and $\chi_W$ using the algorithm discussed in 
\cite{Cha03}, for lattice sizes ranging from $L=32$ to $L=750$ and the 
number of colors from $N=1$ to $N=32$. We vary $T$ with fixed $L_t=4$ 
to study the critical behavior.

\begin{figure}[hbt]
\vskip0.3in
\begin{center}
\includegraphics[width=0.45\textwidth]{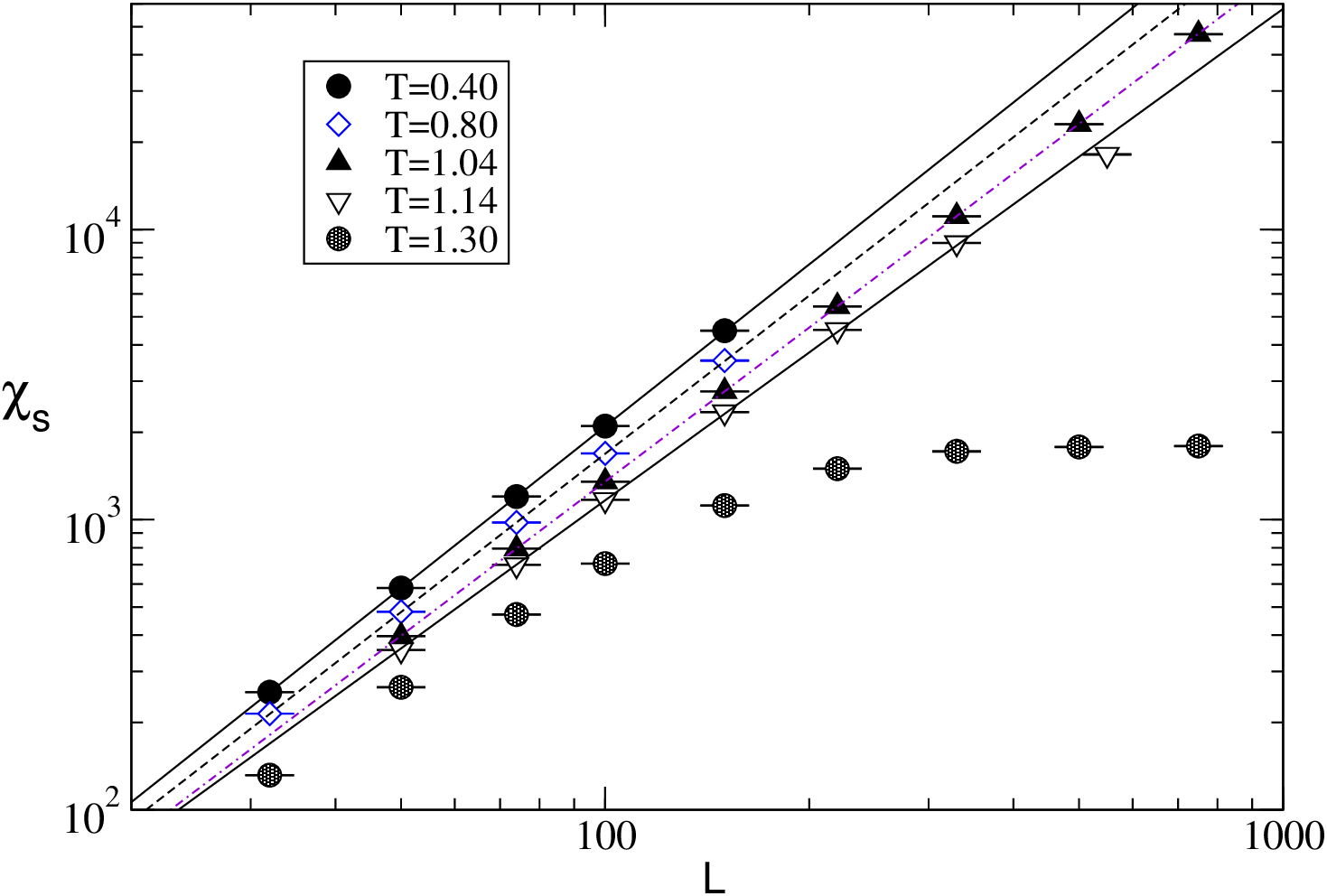}
\end{center}
\caption{\label{fig1}
Plot of $\chi_s$ vs. $L$ for various values of $T$.}
\end{figure}

\begin{figure}[ht]
\vskip0.3in
\begin{center}
\includegraphics[width=0.45\textwidth]{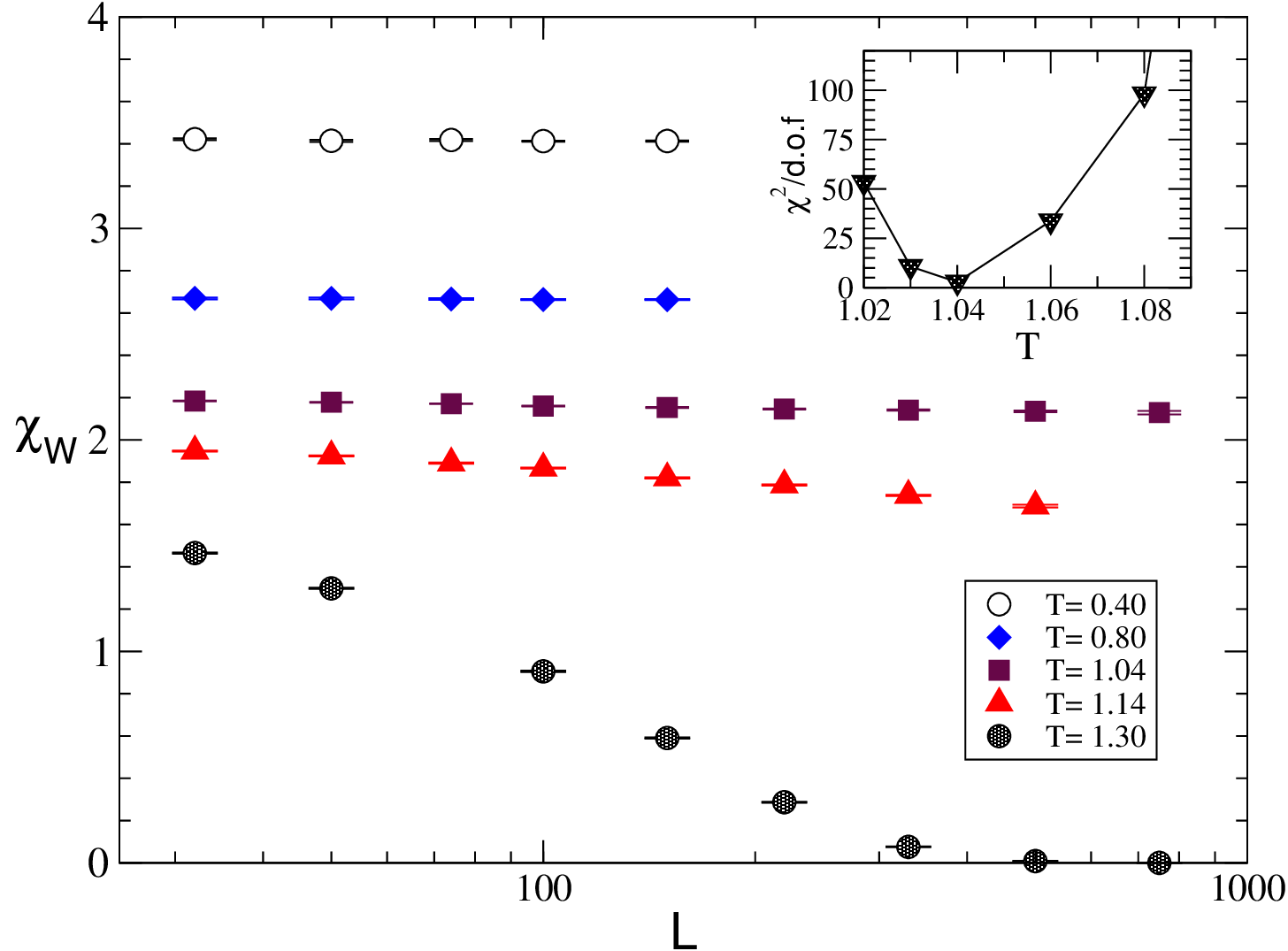}
\end{center}
\caption{\label{fig2}
The plot of $\chi_W$ as a function of $L$ for various
values of $T$. The inset shows $\chi^2$/d.o.f vs. $T$ 
for the fit $\chi_W = 2+1/\log(L/L_0)$ which is valid only at 
$T_c$.}
\end{figure}

Let us first discuss our results for $N=1$. In figures \ref{fig1} and 
\ref{fig2} we plot $\chi_s$ and $\chi_W$ as functions of $L$. We find 
that $\chi_s$  fits extremely well to the form $b L^{2-\eta}$ when 
$T \leq 1.0$ with $\eta$ changing continuously in the expected region. 
In table \ref{tab1} we show the fits along with their $\chi^2$ per degree 
of freedom ($\chi^2/$d.o.f.). In the last column we 
give the results for $1/(2\chi_W)$ on the largest lattices that we could 
compute it. This value matches the value of $\eta$ obtained by the fits. 
The figures and the fits also indicate that when $T \geq 1.10$ one is in the 
high temperature phase with a finite correlation length. 

\begin{table}[ht]
\begin{tabular}{|c|c|c|c|c|}
\hline
$T$ & $b$ & $\eta$ & $\chi^2/$d.o.f & $1/(2\chi_W)$ \\
\hline
0.40 &  0.413(2) &   0.147(1)  &   0.5   &        0.147(1) \\

0.80 &  0.401(2) &   0.188(1)  &    2.4  &        0.189(1) \\

1.00 &  0.396(1) &   0.2236(5) &    0.2  &       0.2256(7) \\

1.02 &  0.397(1) &   0.2293(4) &    1.5  &        0.2321(7) \\

1.04 &  0.400(1) &   0.2356(4) &    2.0  &        0.2378(8) \\

1.06 &  0.402(1) &   0.2421(5) &    0.7  &        0.246(1)  \\

1.10 &  0.413(1) &   0.2604(4) &    6.5   &        0.269(1) \\

1.14 &  0.463(1) &   0.3005(6) &    693  &         0.0000   \\
\hline
\end{tabular}
\caption{\label{tab1} Fits for $\chi_s = b L^{2-\eta}$. The last
column gives the value of $1/2\chi_W$ on the largest lattices.}
\end{table}

When $1.02 \leq T \leq 1.06$, we find that the values of $\eta$ 
do not match $1/2\chi_W$ even on the largest lattices while their
values are close to a quarter as expected near the phase transition.
There are noticeable finite size effects in $\chi_W$. Assuming that 
these discrepancies are due to the logarithmic corrections we fit the 
data for $\chi_s$ to the form $b L^{2-\eta} (\log(L))^{-2r}$. In 
table \ref{tab2} we show these new fits. We also fit the data for 
$\chi_W$ 
to the form $(c + 1/\log(L/L_0))$ following \cite{Har98}. The last column of 
table \ref{tab2} shows the values of $1/2c$ obtained from the fit.
\begin{table}[ht]
\begin{tabular}{|c|c|c|c|c|c|}
\hline
$T$ & $b$ & $\eta$ & 2r  & $\chi^2/$d.o.f & $1/(2 c)$ \\
\hline
1.00 &  0.398(5) &    0.222(5) &  0         & 0.2    &     0.2343(8) \\

1.02 &  0.391(5) &   0.235(5)  &  -0.3(2)   & 1.5     &    0.2411(5) \\

1.04 &  0.384(5) &    0.251(5) &  -0.07(2)  & 0.4     &   0.2483(5) \\

1.06 &   0.395(5)&    0.249(5) &   -0.03(2) & 0.5    &    0.2583(5) \\

1.10 & 0.285(4)  &   0.388(5)  & -0.12(2)   & 3.6     &  0.2831(5)\\

1.14 &  0.243(4) &   0.569(6)  &  -1.24(3)  & 480   &  ---\\
\hline 
\end{tabular}
\caption{\label{tab2} Fits for $\chi_s = b L^{2-\eta}(\log(L))^{-2r}$. The last
column gives the value of $1/2c$ obtained from the fits to $\chi_W$ discussed
in the text.}
\end{table}
We find that the logarithmic terms are unimportant for $T \leq 1.0$
as expected. However, they appear to affect the value of $\eta$ and 
$1/(2c)$ (which is $1/(2\chi_W)$ in the thermodynamic limit) in the range
$1.02 \leq T \leq 1.06$. In particular the value of $2r$ is consistent with 
expectations at $T=1.04$. Further, the new 
values of $\eta$ are closer to $1/2c$. Using the fact that 
$\chi_W = (2 + 1/\log(L/L_0))$ is exactly valid at $T = T_c$ 
we fit the data for $\chi_W$ to this form for various values of $T$. In the 
inset of figure \ref{fig2} we plot $\chi^2$/d.o.f as a function of $T$. 
Based on where the minimum occurs we estimate $T_c=1.040(5)$.

We have checked that $\chi_s$ and $\chi_W$ show similar evidence for
a BKT transition at higher values of $N$. Using techniques similar to 
the $N=1$ we have computed $T_c$ for various values of $N$. We find 
that $T_c = 0.708(6) N + 1.40(4) - 1.07(4)/N$ fits our results very 
well for all values of $N$ with a $\chi^2$/d.o.f of $1.1$. The detailed
analysis of results at higher $N$ and the dependence of non-universal 
critical properties on $L_t$ will be presented elsewhere.

With regards to the $N$ dependence of our results we find two interesting
observations. When $N$ approaches infinity it is often believed that a mean 
field analysis becomes useful. 
\begin{figure}[hbt]
\vskip0.3in
\begin{center}
\includegraphics[width=0.45\textwidth]{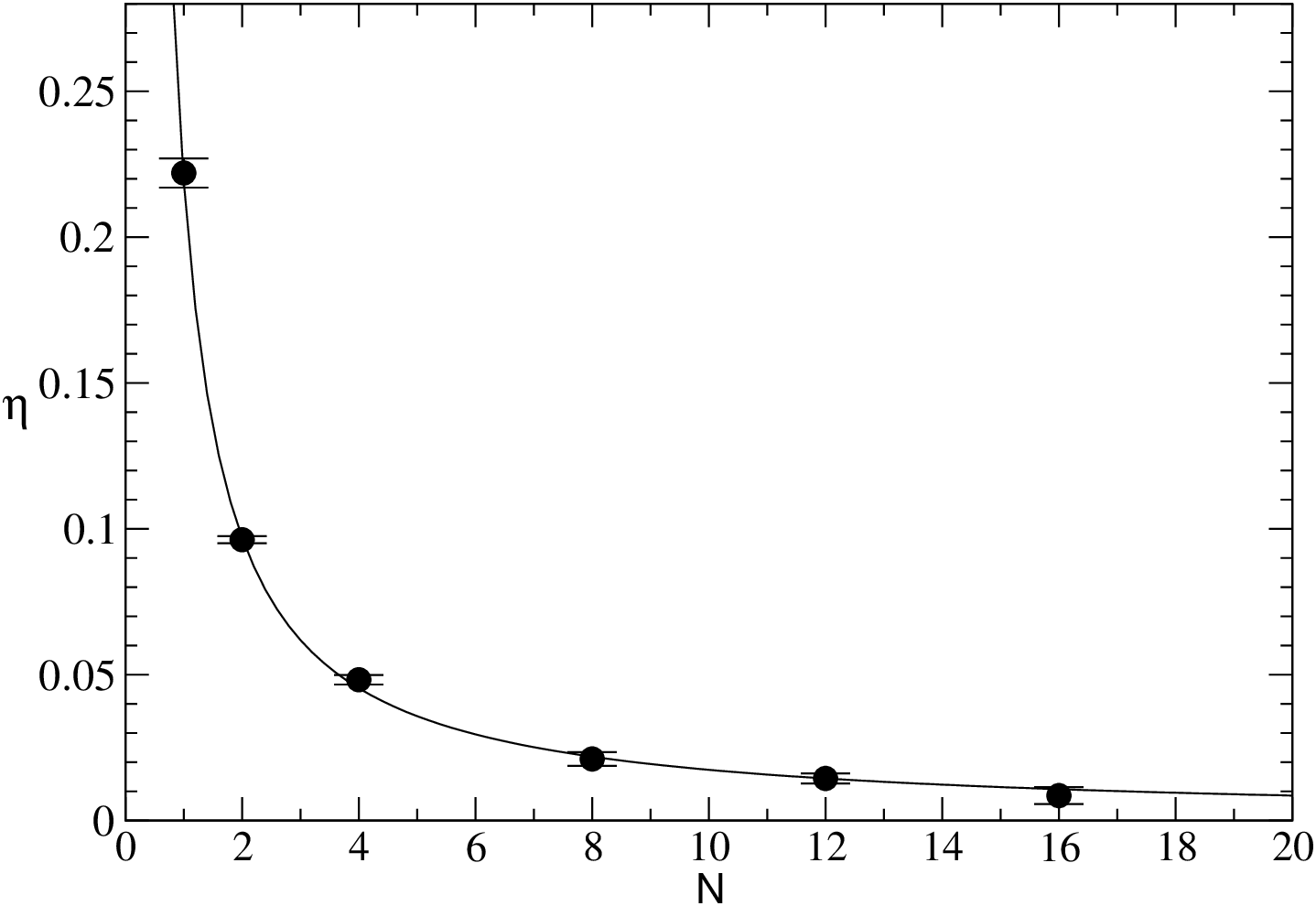}
\end{center}
\caption{\label{fig3}
The plot of $\eta$ as a function of $N$ for $T=1$. The data fits very 
well to the form $\eta = 0.169(6)/N + 0.050(9)/N^2$ with 
$\chi^2/$d.o.f.$=1.2$.}
\end{figure}
Here we argue that there are exceptions to this lore. Consider the 
Mermin-Wagner-Coleman theorem which shows that a continuous symmetry 
cannot break in two dimensions \cite{Mer66}. However, it can be shown that 
$\eta=0$ at $N\rightarrow \infty$ in the mean field approach, which means 
that the condensate is non-zero. Clearly, this cannot be true at any 
large but finite $N$. Witten has argued that when the symmetry is $U(1)$ 
the large $N$ analysis is still applicable since $\eta \sim 1/N$ at
large $N$ \cite{Wit78}. Our results for a fixed $T=1.0$, shown in 
figure \ref{fig3}, do agree with this conjecture.
\begin{figure}[htb]
\vskip0.3in
\begin{center}
\includegraphics[width=0.45\textwidth]{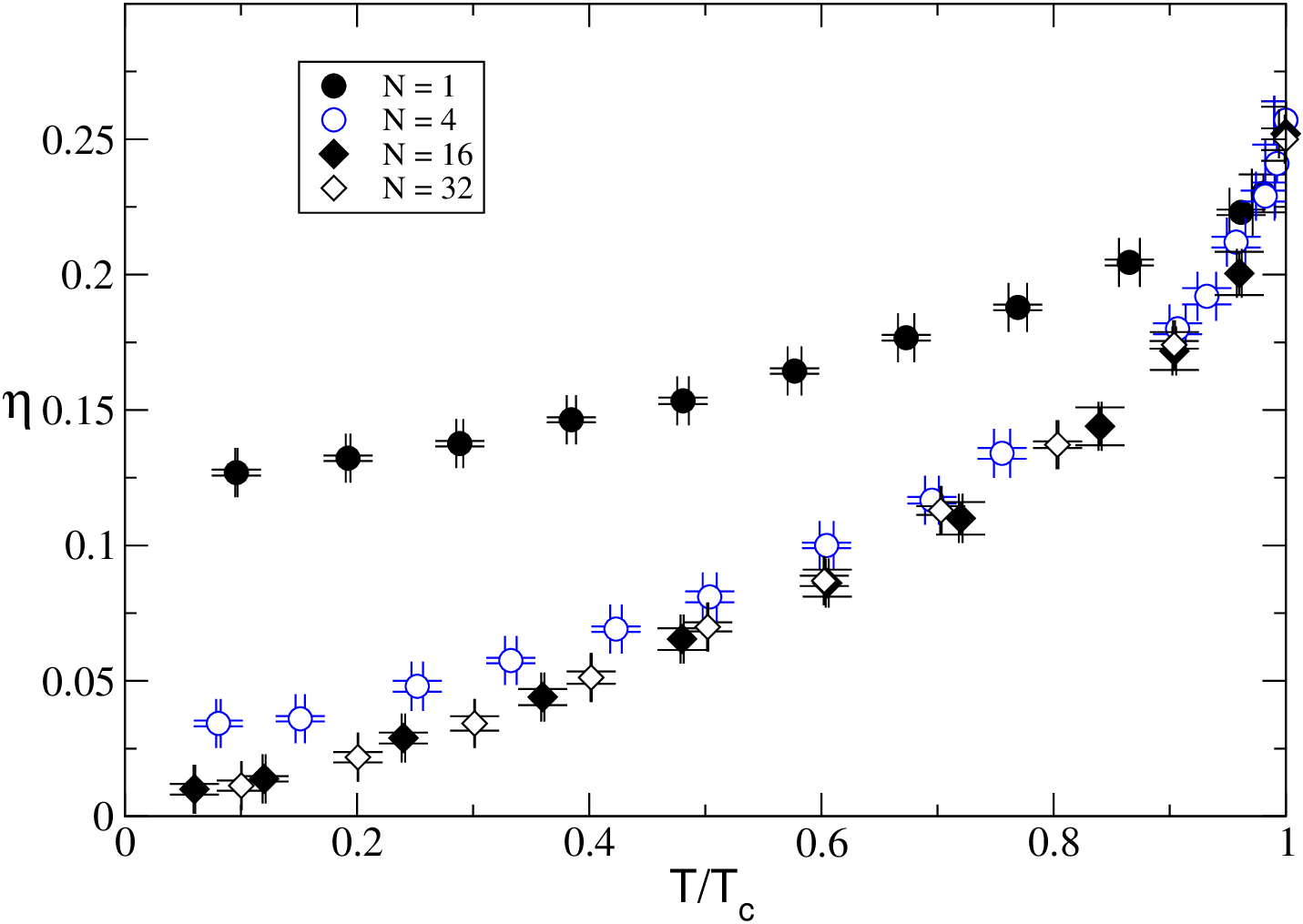}
\end{center}
\caption{\label{fig4}
Plot of $\eta$ as a function of $T/T_c$.}
\end{figure}
Interestingly, as $N$ becomes large and $T/T_c$ is held fixed instead of 
$T$, we find that $\eta \neq 0$ even in the large $N$ limit. Figure 
\ref{fig4} shows that $\eta$ approaches an interesting function of $T/T_c$ 
as $N$ becomes large. Extending this observation to QCD, we think that the 
t'Hooft limit (large N, small gauge coupling $g$, with $g^2 N$ held fixed) 
may be quite similar \cite{tHo75}.

In strictly two dimensions the monomer-monomer correlations for $N=1$ were 
calculated in \cite{Fis63,Har66}. It was found that $\eta \sim 0.5$, which
was recently confirmed again in \cite{Cha03}. This is not in the range 
$0\leq \eta < 0.25$ we found above. We think a closer examination of this 
result in the context of BKT transitions would be interesting.

Finally we note that the $N=1$ model has been recently studied 
in \cite{Hus03} on various types of lattices, using another algorithm 
which is efficient in measuring dimer-dimer correlations at zero monomer 
density \cite{Wer02}. The authors of \cite{Hus03} show the existence of 
a Coulomb phase in three dimensions on cubic lattices. Our results support 
this observation indirectly, however we interpret the results differently. 
We use the connection of the dimer model with strongly coupled lattice QED 
to show that the long range correlations are a result of a spontaneous 
breaking of a global $U(1)$ chiral symmetry in three dimensions. The
winding numbers $W_x$ and $W_y$ we have defined are exactly the magnetic 
flux in the $x$ and the $y$ directions obtained using the magnetic field 
defined in \cite{Hus03}. We interpret the divergence free magnetic field 
as the conserved current related to the $U(1)$ chiral symmetry. 

We would like to thank P. Hasenfratz, R. Moessner, S. Sondhi and U.-J. Wiese
for useful comments. This work was supported in part by the National
Science Foundation grant DMR-0103003 and the U.S. Department of 
Energy grant DE-FG-96ER40945.

\end{document}